\documentclass[a4paper,10pt,twoside]{cpc-hepnp}

\usepackage{multicol}
\usepackage{graphicx}
\usepackage{booktabs}
\usepackage{amssymb,bm,mathrsfs,bbm,amscd}
\usepackage[tbtags]{amsmath}
\usepackage{lastpage}
\usepackage{CJK}

\begin{document}
\begin{CJK*}{GBK}{song}

%\fancyhead[c]{\small Chinese Physics C}
\fancyfoot[C]{\small 010201-\thepage}

\title{\boldmath Measurement of the integrated luminosities of the data taken
  by BESIII at $\sqrt s=$ 3.650 and 3.773 GeV
  \thanks{Supported in part by
the Ministry of Science and Technology of China under Contract No. 2009CB825200;
National Natural Science Foundation of China (NSFC) under Contracts Nos. 10625524,
10821063, 10825524, 10835001, 10935007, 11125525, 11235011; Joint Funds of the
National Natural Science Foundation of China under Contracts Nos. 11079008, 11179007;
the Chinese Academy of Sciences (CAS) Large-Scale Scientific Facility Program; CAS
under Contracts Nos. KJCX2-YW-N29, KJCX2-YW-N45; 100 Talents Program of CAS; German
Research Foundation DFG under Contract No. Collaborative Research Center CRC-1044;
Istituto Nazionale di Fisica Nucleare, Italy; Ministry of Development of Turkey under
Contract No. DPT2006K-120470; U. S. Department of Energy under Contracts Nos. DE-FG02-04ER41291,
DE-FG02-05ER41374, DE-FG02-94ER40823; U.S. National Science Foundation; University of Groningen (RuG)
and the Helmholtzzentrum fuer Schwerionenforschung GmbH (GSI), Darmstadt; WCU Program of National
Research Foundation of Korea under Contract No. R32-2008-000-10155-0.}}

\maketitle

%\author{Author list}
\begin{small}
\begin{center}
M.~Ablikim$^{1}$, M.~N.~Achasov$^{7,a}$, O.~Albayrak$^{4}$,
D.~J.~Ambrose$^{40}$, F.~F.~An$^{1}$, Q.~An$^{41}$, J.~Z.~Bai$^{1}$,
R.~Baldini Ferroli$^{18A}$, Y.~Ban$^{27}$, J.~Becker$^{3}$
% ...
\\
\vspace{0.2cm}
(BESIII Collaboration)\\
\vspace{0.2cm} {\it

$^{1}$ Institute of High Energy Physics, Beijing 100049, People's Republic of China\\
% ...
$^{3}$ Bochum Ruhr-University, D-44780 Bochum, Germany\\
$^{4}$ Carnegie Mellon University, Pittsburgh, Pennsylvania 15213, USA\\
% ...
$^{7}$ G.I. Budker Institute of Nuclear Physics SB RAS (BINP), Novosibirsk 630090, Russia\\
% ...
$^{18}$ (A)INFN Laboratori Nazionali di Frascati, I-00044, Frascati, Italy; (B)INFN and University of Perugia, I-06100, Perugia, Italy\\
% ...
$^{27}$ Peking University, Beijing 100871, People's Republic of China\\
% ...
$^{40}$ University of Rochester, Rochester, New York 14627, USA\\
$^{41}$ University of Science and Technology of China, Hefei 230026, People's Republic of China\\
% ...
$^{a}$ Also at the Novosibirsk State University, Novosibirsk, 630090, Russia\\
% ...
}\end{center}

\vspace{0.4cm}
\end{small}

\begin{abstract}
\small

Data sets were collected with the BESIII detector at the BEPCII
collider at the center-of-mass energy of $\sqrt s =$ 3.650 GeV during
May 2009 and at $\sqrt s=$ 3.773 GeV from January 2010 to May 2011. By
analyzing the large angle Bhabha scattering events, the integrated
luminosities of the two data sets are measured
%at $\sqrt s=$ 3.650 and 3.773 GeV
to be ($\rm 44.49\pm0.02\pm0.44$) $\rm pb^{-1}$ and ($\rm
2916.94\pm0.18\pm29.17$) $\rm pb^{-1}$, respectively, where the first
error is statistical and the second error is systematic.
\end{abstract}



\footnotetext[0]{\hspace*{-3mm}\raisebox{0.3ex}{$\scriptstyle\copyright$}2013
Chinese Physical Society and the Institute of High Energy Physics
of the Chinese Academy of Sciences and the Institute
of Modern Physics of the Chinese Academy of Sciences and IOP Publishing Ltd}%

\begin{multicols}{2}

\section{Introduction}
In $e^+e^-$ collider experiments, the number of events for
$e^+e^-\rightarrow X$ observed in a data set can be written as
\begin{eqnarray}
N^{\rm obs}_{e^+e^-\rightarrow X}(\sqrt s) =
L(\sqrt s) \times
\epsilon_{e^+e^-\rightarrow X}(\sqrt s) \times
\sigma^{\rm obs} (\sqrt s),
\label{eq:chi_TOF}
\end{eqnarray}
where $X$ denotes some final state produced in $e^+e^-$ annihilation,
$N^{\rm obs}_{e^+e^-\rightarrow X}$ is the number of events observed,
$\epsilon_{e^+e^-\rightarrow X}$ is the detection efficiency for
$e^+e^- \rightarrow X$, $L$ is the integrated luminosity and $\sqrt s$
is the center-of-mass energy.

To systematically study the properties of the production and decays of
$\psi(3770)$ and $D$ mesons, a data set was taken at $\sqrt{s} = $
3.773 GeV, with the BESIII detector at the BEPCII, from January 2010
to May 2011.  So far, this data set is the largest $e^+e^-$ collision
data set taken around the $\psi(3770)$ resonance peak in the world.
In order to estimate the continuum contribution in the studies of the
resonance decays, another data set was taken in 2009 at $\sqrt{s} = $
3.650 GeV, which is far away from the resonance peak. The data taken
at $\sqrt{s} = $ 3.773 GeV was accumulated in different periods of
BESIII running; the first part was taken from January 2010 to June
2010 and the second part was taken from December 2010 to May 2011. For
convenience in the following, we call the data taken at $\sqrt{s} = $
3.650 GeV as the continuum data and call the two parts of the data
taken at $\sqrt{s} = $ 3.773 GeV as $\psi(3770)$ data A and
$\psi(3770)$ data B, respectively.

In this paper, we present the measurements of the intgrated
luminosities of the data sets taken at $\sqrt{s}$ = 3.650 and 3.773
GeV by analyzing the large angle Bhabha scattering events.

\section{BESIII detector}

The BESIII detector and the BEPCII collider \cite{bes3} are  major
upgrades of the BESII detector and the BEPC collider \cite{bes2}. The
design peak luminosity of the double-ring $e^+e^-$ collider, BEPCII,
is $10^{33}$ cm$^{-2}s^{-1}$ at a beam current of 0.93 A. The peak
luminosity at $\sqrt{s} = $ 3.773 GeV reached 0.65 $\times
10^{33}$ cm$^{-2}s^{-1}$ in April 2011 during the $\psi(3770)$ data
taking. The BESIII detector with a geometrical acceptance of 93\% of
4$\pi$, consists of the following main components: 1) a
small-celled, helium-based main draft chamber (MDC) with 43 layers.
%equipped with 6796 signal wires and 21884 field wires arranged in a small cell
%configuration with 43 layers working in a gas mixture of He (40\%) and
%C$_3$H$_8$ (60\%).
The average single wire resolution is 135 $\rm \mu m$, and the
momentum resolution for 1~GeV/$c$ charged particles in a 1 T magnetic
field is 0.5\%; 2) an electromagnetic calorimeter (EMC) made of 6240
CsI (Tl) crystals arranged in a cylindrical shape (barrel) plus two
endcaps. For 1.0 GeV photons, the energy resolution is 2.5\% in the
barrel and 5\% in the endcaps, and the position resolution is 6 mm in
the barrel and 9 mm in the endcaps; 3) a Time-Of-Flight system (TOF)
for particle identification composed of a barrel part made of two
layers with 88 pieces of 5 cm thick, 2.4 m long plastic scintillator
in each layer, and two endcaps with 96 fan-shaped, 5 cm thick, plastic
scintillators in each endcap. The time resolution is 80~ps in the
barrel, and 110 ps in the endcaps, corresponding to a $2\sigma$
K/$\pi$ separation for momenta up to about 1.0 GeV/$c$; 4) a muon chamber
system (MUC) made of 1600 m$^2$ of Resistive Plate Chambers (RPC)
arranged in 9 layers in the barrel and 8 layers in the endcaps and
incorporated in the return iron of the superconducting magnet. The
position resolution is about 2 cm.

\section{Method}

In principle, any QED process can be used to measure the
integrated luminosity of the data set using
\begin{eqnarray}
L(\sqrt s)= \frac {N^{\rm obs}_{\rm QED}(\sqrt s) \times
(1-\eta)}{\sigma_{\rm QED}(\sqrt s) \times \epsilon \times
\epsilon^{\rm trig}_{e^+e^-}} , \label{eq:cal_lum}
\end{eqnarray}
where $\rm N^{obs}_{QED}$ is the observed number of events of the final
state in question, $\rm \sigma_{QED}$ is the production
cross section, which can be
determined by theoretical calculation, $\epsilon$ is the detection
efficiency, $\eta$ is the contamination ratio and $\epsilon^{\rm
trig}_{e^+e^-}$ is the trigger efficiency for collecting the QED process in
the on-line data acquisition.

Usually, the processes of $e^+e^-\to (\gamma)e^+e^-$,
$e^+e^-\rightarrow (\gamma)\gamma\gamma$ and $e^+e^-\to
(\gamma)\mu^+\mu^-$ are used to measure the integrated luminosity of
the data because of their simpler final state topologies, larger
production cross sections, higher detection efficiencies, as well as
more precisely expected cross sections from theory.
In this work, the large angle Bhabha scattering events of $e^+e^-
\rightarrow (\gamma)e^+e^- $ are adopted.
Throughout the paper, the symbol of ``($\gamma$)" denotes the possible
photon(s) produced due to Initial State Radiation or Final State Radiation.

\section{Luminosity measurement}

\subsection{Event selection}

In order to select candidate Bhabha events, it is required that
there should be only two good charged tracks with total charge zero,
which are reconstructed in the MDC. Each track must originate from the
interaction region of $\rm R_{xy} <$ 1 cm and $\rm |V_z| < $ 5 cm,
where $\rm R_{xy}$ and $\rm |V_z|$ are the points of closest
approach relative to the collision point in the xy-plane and in the z
direction, respectively. Furthermore, to ensure that the candidate
charged track hits the barrel of the EMC, we require that the polar
angle $\theta$ of the charged track satisfy $\rm |\cos \theta| <$ 0.80.

\begin{center}
\centering
\includegraphics[width=8cm,height=4.5cm]{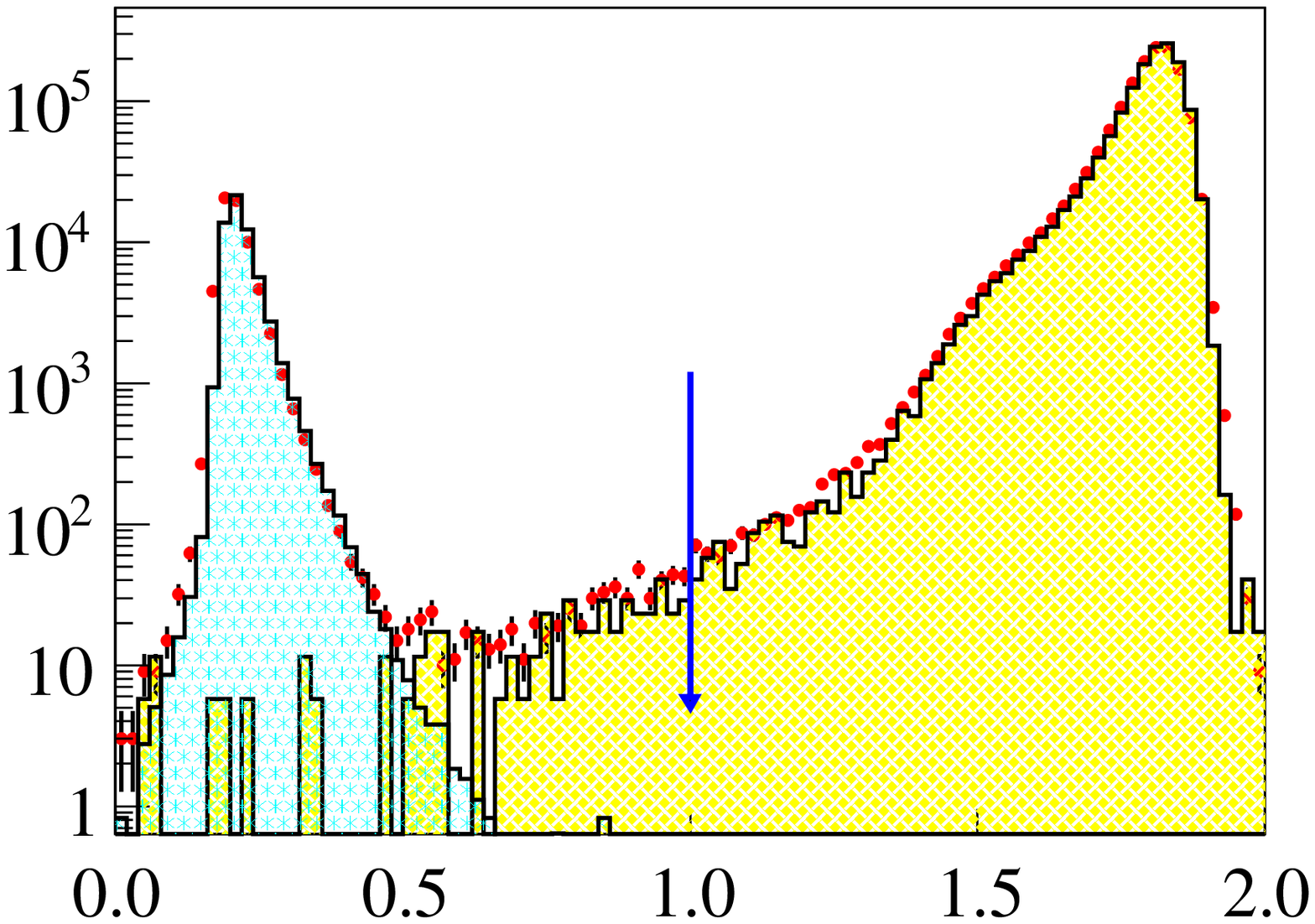}
\put(-145,0){\normalsize \bf $\rm E_{EMC}$(GeV)} \put(-220,20)
{\normalsize \rotatebox{90} {\bf Events(/0.01 GeV)}} \figcaption{
\footnotesize The distributions of the deposited energies in the EMC
of the charged tracks from the selected events, where the dots with
red error bars are the continuum data, the yellow
histogram is the Monte Carlo events of $e^+e^- \rightarrow
(\gamma)e^+e^-$ and the light green histogram is the
Monte Carlo events of $e^+e^- \rightarrow (\gamma)\mu^+\mu^-$.}
\label{fig:Eraw}
\end{center}

Figure \ref{fig:Eraw} shows the deposited energies in the EMC ($\rm
E_{EMC}$) for the good charged tracks of events satisfying the above
selection criteria, where the dots with red error bars are the
continuum data, the yellow histogram is $e^+e^- \rightarrow
(\gamma)e^+e^-$ Monte Carlo events and the light green histogram is
$e^+e^- \rightarrow (\gamma)\mu^+\mu^-$ Monte Carlo events. From the
figure it can be seen that the requirement $\rm E_{EMC} > 1.0$ GeV can
cleanly separate the $e^+e^-\rightarrow (\gamma)\mu^+\mu^-$ events
from the Bhabha scattering events.  To further remove background from cosmic
rays, the momenta of the two good charged tracks in the candidate
Bhabha events should not both be greater than $\rm E_b + 0.15$ GeV,
where $\rm E_{b}$ is the calibrated beam energy.

After applying the above selection criteria, the accepted events are
mostly Bhabha scattering events. But there may be still a small amount
of background from $e^+e^- \rightarrow (\gamma)J/\psi$, $e^+e^-
\rightarrow (\gamma) \psi(3686) \rightarrow (\gamma)J/\psi X $ and
$e^+e^- \rightarrow \psi(3770) \rightarrow (\gamma) J/\psi X$ ($J/\psi
\to e^+e^-$ and $\rm X = \pi^0\pi^0, \eta, \pi^0$ or $\gamma\gamma $).
In order to remove these background events, the sum of the momenta of
the two good charged tracks is required to be greater than $\rm
0.9\times E_{cm}$. The remaining contamination from these background
sources are estimated by Monte Carlo simulation, which will be
discussed in Section 4.3.

\subsection{Data analysis}

The two oppositely charged tracks in the candidate Bhabha scattering
events are bent in the magnetic field, so the positions of their two
shower clusters in the xy-plane of the EMC are not back-to-back.  To
determine the observed number of Bhabha scattering events, we use the
difference of the azimuthal angles of the two clusters in the EMC,
which is defined as $\delta \phi$ = $|\phi_1 - \phi_2|$-180 in
degrees, where $\phi_1$ and $\phi_2$ are the azimuthal angles of the
two clusters in the EMC.  Figure \ref{fig:delphi} shows the $\delta
\phi$ distribution of the candidate Bhabha scattering events selected
from the continuum data. In the figure, the events in the ``signal"
regions between the red arrows are taken as the signal events, while
the ones in the ``sideband" regions between the blue arrows are used
to estimate the background in the $\delta \phi$ ``signal"
region. After subtracting the scaled number of the events in the
sideband region from the number of events in the signal region, we
obtain the numbers of the Bhabha scattering events observed from
data, which are listed in the second row of Table
\ref{Tab:results_lum}.

\begin{center}
\includegraphics[width=8cm,height=4.5cm]{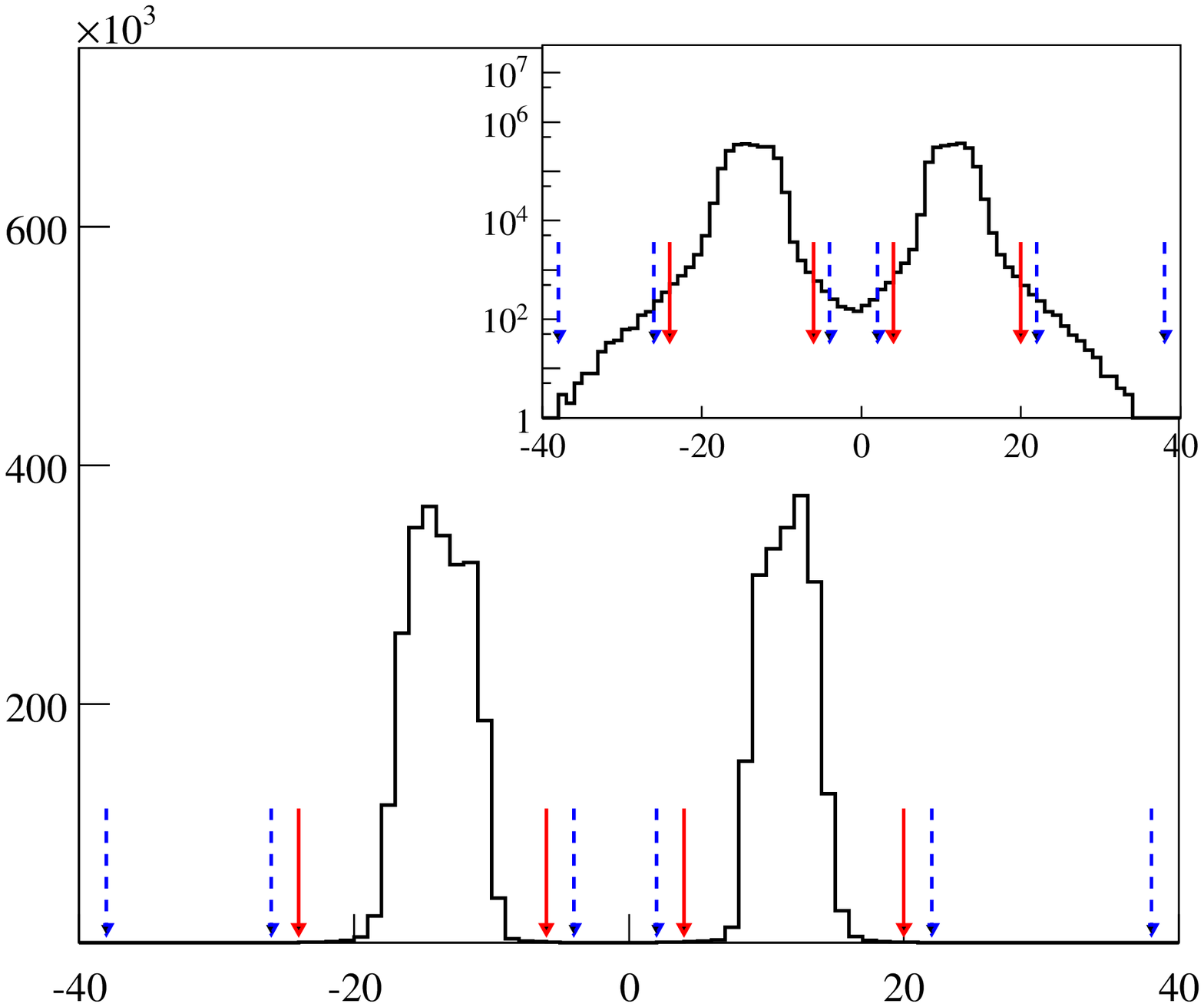}
\put(-140,0){\normalsize \bf $\delta \phi$ in degrees}
\put(-230,30){\normalsize \rotatebox{90} {\bf Events(/degree)}}
 \figcaption{\label{fig:delphi}  The
distribution of $\delta \phi$ ($\delta \phi$ =  $|\phi_1 -
\phi_2|$-180$^{\circ}$) of the selected $e^+$ and $e^-$ tracks.}
\end{center}

\subsection{Background estimation}

For the accepted Bhabha scattering events, there may still be some
residual background from $e^+e^- \rightarrow (\gamma)J/\psi $, $e^+e^-
\rightarrow (\gamma) \psi(3686) \rightarrow (\gamma)J/\psi X $ and
$e^+e^- \rightarrow \psi(3770) \rightarrow (\gamma) J/\psi X$ ($J/\psi
\to e^+e^-$ and $\rm X = \pi^0\pi^0, \eta, \pi^0$ or $\gamma\gamma $),
as well as some other hadronic decay processes. These are estimated by
analyzing the Monte Carlo events, including 16.5 M $e^+e^- \rightarrow
(\gamma)J/\psi$, 51 M $e^+e^- \rightarrow (\gamma) \psi(3686)$, 198 M
$e^+e^- \rightarrow \psi(3770) \to D\bar{D}$, 15 M $e^+e^- \rightarrow
\psi(3770) \to$ non-$D\bar{D}$, and 183 M $e^+e^- \rightarrow $
continuum light hadron events. Detailed analysis gives the
contamination rates to be $\eta = 1.7 \times 10^{-5}$ and $1.7 \times
10^{-4}$ for the candidate Bhabha scattering events selected from the
continuum data and the $\psi(3770)$ data, respectively.

\subsection{\boldmath Detection efficiency for $e^+e^-\rightarrow (\gamma)e^+e^-$}

To determine the detection efficiencies for the Bhabha scattering
events, we generate 400,000 $e^+e^- \rightarrow (\gamma)e^+e^-$ Monte
Carlo events with the Babayaga generator \cite{generator} within the
polar angle range of $|\cos \theta| <$ 0.83 at $\sqrt{s} = $ 3.650 and 3.773
GeV, where $\theta$ is the polar angle of the $e^+$ and $e^-$. By
analyzing these Monte Carlo events with the same selection criteria as
the data analysis, we obtain the detection efficiencies for $e^+e^-
\rightarrow (\gamma)e^+e^-$ at $\sqrt{s} = $ 3.650 and 3.773 GeV,
which are summarized in the fourth row of Table \ref{Tab:results_lum}.

\subsection{Integrated luminosities}

Inserting the numbers of observed Bhabha scattering events, the
detection efficiencies for $e^+e^-\rightarrow (\gamma)e^+e^-$ obtained
by the Monte Carlo simulation, the trigger efficiency and the visible
cross sections within the polar angle range of $|\cos \theta| < $
0.83 in Eq. (\ref{eq:cal_lum}), we determine the integrated
luminosities of the continuum data, the $\psi(3770)$ data A and the
$\psi(3770)$ data B to be ($\rm 44.49\pm0.02\pm0.44$) $\rm pb^{-1}$,
($\rm 927.67\pm0.10\pm9.28$) $\rm pb^{-1}$ and ($\rm
1989.27\pm0.15\pm19.89$) $\rm pb^{-1}$, respectively, where the first
errors are statistical and the second are systematic and discussed in
the next section.  The total luminosity of the $\psi(3770)$ data is
($\rm 2916.94\pm0.18\pm29.17$) $\rm pb^{-1}$.  Here, for the data sets
used in the analysis, the trigger efficiency for collecting
$e^+e^-\rightarrow (\gamma)e^+e^-$ events was determined to be
$\epsilon^{\rm trig}_{e^+e^-} = $ 100$\%$ with the statistical error
being less than 0.1$\%$~\cite{trig}.  The numbers used in the
luminosity measurements are summarized in Table \ref{Tab:results_lum}.

\end{multicols}

\begin{center}
\tabcaption{ Summary of the numbers used in the determination of the
luminosities, where $N^{\rm obs}_{e^+e^-\rightarrow (\gamma)e^+e^-}$
is the number of candidate Bhabha scattering events selected from the
data, $\epsilon$ is the detection efficiency, $\sigma$ is the visible
cross section for the Bhabha scattering events and $L$ represents the
integrated luminosity.}
\vspace{-3mm} \footnotesize
\begin{tabular*}{170mm}{@{\extracolsep{\fill}}lcccccc}
\toprule
Samples                                            & $\psi(3770)$ data A  &  $\psi(3770)$ data B & continuum data  \\
\hline
$N^{\rm obs}_{e^+e^-\rightarrow (\gamma)e^+e^-}$ ($\times 10^{4}$)  & $8412.9\pm 0.9$  & $18140.3\pm1.3$ & $432.0\pm0.2$  \\
$\eta$ ($\times 10^{-4}$)                          & 1.7            & 1.7              & 0.17         \\
$\epsilon$ (\%)                                    & 61.28          & 61.62            & 61.47         \\
$\sigma$ [nb]                                      & 147.9599       & 147.9599         & 157.9393      \\
\hline
$L$ [$\rm pb^{-1}$]                                  & $\rm 927.67\pm0.10\pm9.28$        & $\rm 1989.27\pm0.15\pm19.89$                      & $\rm 44.49\pm0.02\pm0.44$   \\
\bottomrule \label{Tab:results_lum}
\end{tabular*}
\end{center}

\begin{multicols}{2}

\subsection{Systematic error}

In the measurements of the integrated luminosities, the systematic
errors arise from the uncertainties associated with the Bhabha event
selection, the Monte Carlo statistics, the background estimation, the
signal region selection, the trigger efficiency and the generator.

In order to estimate the systematic uncertainty due to the $\cos
\theta$ requirement, we also determine the integrated luminosities
with the selection requirements of $|\cos \theta |<$ 0.75 and 0.70,
and the differences from the standard selection of $|\cos \theta |<$
0.80 are all less than 0.5\% for both the continuum data and
$\psi(3770)$ data. To be conservative, we take 0.75\% as the
systematic error due to the $\cos \theta$ selection in this work.  The
systematic uncertainty due to the MDC measurement information, which
includes the uncertainties due to the MDC tracking efficiency and the
momentum requirement, is determined to be 0.3\% by comparing the
integrated luminosities measured with and without the MDC measurement
information.  The systematic uncertainty due to the $E_{EMC}$ energy
selection requirements is determined to be 0.2\%, by comparing the
$E_{EMC}$ distributions of the data and Monte Carlo events.  The
uncertainty from the $E_{EMC}$ cluster reconstruction is determined to be
0.03\% by comparing the efficiencies of the data and the Monte Carlo
events.

The uncertainty from the Monte Carlo statistics is 0.1\%.  The
uncertainty in the background subtraction is negligible.  The
uncertainty due to the $\delta \phi$ signal region selection is
estimated to be 0.01\% by comparing the integrated luminosities
measured with different signal regions. In these measurements, we use
the trigger efficiency for collecting $e^+e^-\rightarrow
(\gamma)e^+e^-$ events of $\epsilon^{\rm trig}_{e^+e^-} = $ 100$\%$
with the statistical error being less than 0.1\% \cite{trig}.
Therefore, we take 0.1\% as the systematic uncertainty due to trigger
efficiency. The uncertainty due to the Bhabha generator is 0.5\%,
which is cited from Ref. \cite{generator}.

\begin{center}
\tabcaption{The relative systematic uncertainties in the luminosity
measurement.} \footnotesize
\begin{tabular*}{80mm}{l@{\extracolsep{\fill}}ccc}
\toprule Sources & $\Delta^{\rm sys}$ ($\%$)  \\
\hline
$|\cos \theta| <$ 0.80    & 0.75  \\
$\rm E^{e^+}_{EMC}>$ 1 GeV  & 0.2  \\
$\rm E^{e^-}_{EMC}>$ 1 GeV  & 0.2  \\
MDC information             & 0.3  \\
EMC cluster reconstruction    & 0.03 \\
Monte Carlo statistics      & 0.1  \\
Background estimation       & 0.0  \\
Signal region selection ($\delta \phi$)     & 0.01 \\
Trigger efficiency \cite{trig} & 0.1 \\
Generator \cite{generator}  & 0.5  \\
\hline
Total                  & 1.0   \\
\bottomrule \label{Tab:err_lum}
\end{tabular*}
\end{center}

Table \ref{Tab:err_lum} summarizes the above systematic uncertainties
in the luminosity measurement. The total systematic error is
determined to be 1.0$\%$ by adding these uncertainties in quadrature.

\section{Summary}

By analyzing the Bhabha scattering events, we measure the
integrated luminosities of the data taken at $\sqrt{s} = $ 3.650 and
3.773 GeV with the BESIII detector to be ($\rm 44.49\pm0.02\pm0.44$)
$\rm pb^{-1}$ and ($\rm 2916.94\pm0.18\pm29.17$) $\rm pb^{-1}$,
respectively. These luminosities can be used in normalization in
studies of $\psi(3770)$ production and decays, as well as in studies
of $D$ meson production and decays.

\acknowledgments{The BESIII collaboration thanks the staff of BEPCII and the
computing center for their strong support.}

\end{multicols}

\vspace{-1mm}
\centerline{\rule{80mm}{0.1pt}}
\vspace{2mm}

\begin{multicols}{2}

\end{multicols}

\clearpage

\end{CJK*}

\begin{thebibliography}{90}

\vspace{3mm}


\bibitem{bes3}
BESIII Collaboration, M. Ablikim {\it et al.}, Design and
construction of the BESIII detector, Nucl. Instrum. Meth. {\bf A
614} (2010) 345.

\bibitem{bes2}
BES Collaboration, J. Z. Bai {\it et al.}, Nucl. Instrum. Meth. {\bf
A 344} (1994) 319; Nucl. Instrum. Meth. {\bf A 458} (2001) 627.

\bibitem{trig} Trigger efficiencies at BES-III, N.Berger, Zhu Kai {\it et al.} Chinese Physics {\bf C 34(12)} (2010) 1779-1784.


\bibitem{generator}
C.M. Carloni Calame, G. Montagna, O. Nicrosini, F. Piccinini,
Nucl. Phys. Proc. Suppl. {\bf 131} (2004) 48-55.


\end{thebibliography}
\end{document}